# Observation of unconventional six-fold, four-fold and three-fold excitations in rare-earth-metal carbide Re$_2$C$_3$

Lei Jin,[1,2] Ying Liu,[1,2*] Xiaoming Zhang,[1,2] Xuefang Dai,[1,2] and Guodong Liu[1,2*]

[1]*State Key Laboratory of Reliability and Intelligence of Electrical Equipment, Hebei University of Technology, Tianjin 300130, China*

[2]*School of Materials Science and Engineering, Hebei University of Technology, Tianjin 300130, China.*

E-mail: ying_liu@hebut.edu.cn; gdliu1978@126.com

**ABSTRACT:**

Unconventional fermions, such as three-fold, four-fold, six-fold, and eight-fold fermions have attracted intense attention in recent years. However, the concrete materials hosting unconventional fermions are still in urgent scarcity. In this work, based first-principle calculations and symmetry analysis, we reveal rich unconventional fermions in existing compound Re2C3 (Re = Y, La, Ce, Pr, Nd, Sm, Tb, Dy, Ho, Er, Tm, Yb, Lu). We show that these compounds host quadratic dispersive three-fold (TP), linear dispersive four-fold (FP) and six-fold points (SP) near the Fermi level in their electric band structures when spin-orbital coupling (SOC) is not included. Notably, the FP is charge-2 Dirac-like point. More importantly, among compound Re2C3, the compound Yb2C3 has very clean band structure, and its unconventional fermions are closed to the Fermi level. We also find that a uniaxial strain can transform the unconventional fermions into other types fermions, depending on the directions of strain. When SOC is considered, a SP transform to an eightfold degenerate point and a fourfold degenerate point. Overall, our work provides a family of realistic materials to study the unconventional fermions.

## I. INTRODUCTION

Topological semimetals/metals have been attracting intense research interest in nowadays condensed matter physics [1]. Commonly, there emergence nontrivial band-crossings in their electric band structures near the Fermi level, enforced by topology and symmetry. Hence, the quasiparticles in the low-energy region behave differently from the conventional fermions which are described by the Schrödinger equation. For example, Weyl/Dirac semimetals/metals possess twofold/fourfold degenerate band crossings in their band structures, and the quasiparticle around the crossing can be described by Weyl/Dirac equation which behaves like the Weyl/Dirac fermion in high energy region, [2-20] leading to novel physical phenomena, like the chiral anomaly [12]. Recently, topological semimetals/metals with unconventional fermions, like double Weyl fermion, charge-2 Dirac fermion, spin-1 fermion, six-fold fermion, eightfold fermion, protected by crystalline symmetry are becoming a focusing research field. [21-28] Therefore, it is natural to search the realistic materials to study these unconventional fermions which are accompanied by fascinating physical properties.

Until now, only several materials have been reported that host new types of unconventional fermions. For example, structurally-chiral topological semimetals, CoSi family, [29-33] was predicted to host Chiral fermions that characterized by large Chern numbers $\mathcal{C}$ (topological charges), namely, spin-1 excitation and double Weyl fermions which have $|\mathcal{C}| = 2$. Since these fermions locate at the center or corner of the first Brillouin zone, Fermi arcs connecting the nonzero chiral fermions almost span the whole Brillouin zone [33]. Remarkably, due to the absence of mirror symmetries, these two Chiral fermions locate at different energies, expected to display some exotic physical effects, gyrotropic magnetic effect, quantized circular photogalvanic effect [34,35]. Recently, Takane *et al.*[30] have experimentally demonstrated the existence of these unconventional chiral fermions. Subsequently, a chiral fermion of six-fold degeneracy characterized by $\mathcal{C} = \pm 4$ was found in AlPt material verified in experimental and theoretical works. [36] Followingly, PdSb$_2$ [37-39], PtBi$_2$ [40], and Li$_{12}$Mg$_3$Si$_4$ [41] are also predicted to possess such a fermion. A few real materials, including compound TaTe$_4$ [42], were predicted to host an eightfold fermion. However, the number of candidates to study the unconventional fermions is limited, and they also have their own disadvantages.

In this work, based on the first-principle calculations and symmetry analysis, we reveal a family of materials, Re$_2$C$_3$, exhibiting three types of fermions: a six-fold point at $H$ point, a three-fold point at $\Gamma$ point, and a charge-2 Dirac point at $P$ point. We take Yb$_2$C$_3$ as an example to support our discovery. Particularly, the experimental synthetic Yb$_2$C$_3$ has a clean band structure, easy to observe that there are three unconventional band crossings near the Fermi level. We also construct the low-energy effective $k \cdot p$ model for the three points. We find that the threefold point has a quadratic dispersion which is dramatically different from the spin-1 fermion, and Dirac-like point is a Chiral fermion with a topological charge equal to $\pm 2$. The main features of them can also be read off from the effective Hamiltonian. We discovery that the six-fold degenerate point has a zero Chern number which is a composition of two spin-1 fermions of opposite chirality, and we demonstrate it by the symmetry constrained effective Hamiltonian. Notably, the Fermi arcs span the whole surface Brillouin zone (BZ), which is promising to be directly observed by angle-resolved photoemission spectroscopy (ARPES). Therefore, the Re$_2$C$_3$ family compounds are excellent candidates to investigate the unconventional fermions.

## II. COMPUTATIONAL METHODS

We perform first principles calculations in the framework of DFT. All calculation in our work were used the Vienna ab initio simulation package (VASP) [43]. Ion cores adopt the projector augmented wave pseudopotentials [44] to model. For the exchange correlation-potential, we adopt the generalized gradient approximation (GGA) of the Perdew–Burke–Ernzerhof (PBE) functional [45]. The plane wave basis energy cutoff was set as 500 eV. A $\Gamma$-centered $k$-mesh of $11 \times 11 \times 11$ was used for BZ sampling. For the rare earth elements, we executed GGA + U calculations to describe the Coulomb interaction. [46] The effective Coulomb energy $U_{\text{eff}}$ was set as 3 eV, and we found the band structures will not change in a large range of $U$ values (0−5 eV) [see

Figure S1 in the Supporting Information]. The topological features of surface states were calculated based on the maximally localized Wannier functions [47,48], realized by using the WANNIERTOOLS package [49].

## III. CRYSTAL STRUCTURE

The $Re_2C_3$ (Re = Y, La, Ce, Pr, Nd, Sm, Tb, Dy, Ho, Er, Tm, Yb, Lu) materials have been synthesized in experiments [50-54]. We take $Yb_2C_3$ as an example to illustrate our results. $Yb_2C_3$ has a cubic crystal structure belonging to space group $I\bar{4}3d$ (No. 220). As shown in Fig. 1(a), one Yb atom bonds with six C atoms. In a unit cell, Yb and C atoms occupy the 16$c$ (0.0499, 0.0499, 0.0499) and 24$d$ (0.2937, 0, 0.25) Wyckoff sites, respectively. The fully relaxed lattice constants are a = b =c = 8.348 Å, which match well with the experimental values (a = b = c = 8.072 Å [50]). Figure 1(b) shows the primitive cell of $Yb_2C_3$ with 8 Yb and 12 C atoms. The optimized and experimental lattices of all materials in $Re_2C_3$ family compounds have been summarized in Table 1. In our calculations, we adopt the optimized lattice structures.

## IV. TOPOLOGICAL BAND STRUCTURE

### A. Symmetry analysis and $k \cdot p$ model

We plot the total density of states (DOS) and the projected density of states (PDOS) of the $Yb_2C_3$ compound in the absence of spin-orbit coupling (SOC) in Figs. 1(c) and 1(d). One can observe that the bands around the Fermi level are mainly contributed by the $d$ orbitals from Yb atoms. Figure 2(b) shows the band structure of $Yb_2C_3$ without SOC along high symmetry paths. One can notice that there exist three band crossings, namely, a six-fold point at $H$ point, a three-fold point at $\Gamma$ point, and a four-fold point at $P$ point. Importantly, these multifold-degenerate points are very closed to the Fermi level without other extraneous bands. Thus, they are promising for experimental observations in the near future. Also, the materials show no magnetic ordering, thus the time-reversal symmetry ($T$) is preserved. Considering the nonsymmorphic crystalline symmetries, they together protect these unconventional excitations.

Let us first study the six-fold band crossing at $H$ point. The six-fold degenerate point is near to the Fermi level (locating at 0.194 eV). It is formed by four electronlike bands and two holelike bands, consistent with the 3D band structure that one observes in Fig. 2(c). Along the $\Gamma$–$H$ line in Fig. 2(b), the six-fold band split into four singly-degenerate bands and one doubly-degenerate band. Symmetry analysis shows that five bands belong to irreducible representations $\Gamma_3$, $\Gamma_3$, $\Gamma_1$, $\Gamma_2+\Gamma_4$ and $\Gamma_1$ of the $C_{2v}$ symmetry. In the other path, $H$-$N$ line, the six-fold degenerate bands split into three two-fold degenerate bands protected by $\widetilde{M}_{1\text{-}10}$ and $T$ symmetries. They have same irreducible representations $\Gamma_1+\Gamma_2$ of the $C_s$ symmetry.

One has checked that the topological charge of this six-fold point is zero which can be demonstrated by the effective Hamiltonian. The little group of it belong to $T_d$ which is generated by $\{S_{4x}^{-1}|\frac{1}{2}00\}$, $\{\widetilde{M}_{110}|\frac{1}{2}00\}$ and $\{C_{3,1\overline{1}\overline{1}}^{-1}|1\frac{1}{2}\frac{1}{2}\}$. The irreducible representation of the six-fold band at H is the superposition of Γ4 and Γ5. To characterize the topology of the six-fold excitation, we have constructed an effective k · p model around this band crossing at H. Taking the $\Gamma_4$ and $\Gamma_5$ irreducible representations as the basis, after a unitary transformation, the effective Hamiltonian takes the form as following,

$$\mathcal{H}(\boldsymbol{k}) = \begin{pmatrix} 0 & -ivk_x & -ivk_y & 0 & -v'k_x & v''k_y \\ ivk_x & 0 & ivk_z & -v''k_x & 0 & v'k_z \\ ivk_y & -ivk_z & 0 & v'k_y & v''k_z & 0 \\ 0 & -v''k_x & v'k_y & 0 & ivk_x & ivk_y \\ -v'k_x & 0 & v''k_z & -ivk_x & 0 & -ivk_z \\ v''k_x & v'k_z & 0 & -ivk_y & ivk_z & 0 \end{pmatrix}. \quad (1)$$

We should point out that, in a limitation that $v \gg v'$, and $v \gg v''$, the Hamiltonian can be rewritten as

$$\mathcal{H}(\boldsymbol{k}) = \begin{pmatrix} v\boldsymbol{k} \cdot \boldsymbol{S} & 0 \\ 0 & -v\boldsymbol{k} \cdot \boldsymbol{S} \end{pmatrix}. \quad (2)$$

Here, $\boldsymbol{S}$ is the angular momentum for spin-1 fermion, which satisfies that $[S_i, S_j] = i\epsilon_{ijk}S_k$. Consequently, this six-fold point is a composition of two spin-1 fermions of opposite helicity. As the conventional Dirac point, it also shows a zero topological charge.

Turn to the three-fold band crossing at $\Gamma$ point. Figure 2(b) shows the electronic structure for it along the *N-Γ-P* path without SOC. Along the *N-Γ* line in Fig. 2(b), the three-fold band split into three singly-degenerate bands, which belong to irreducible representations $\Gamma_2$, $\Gamma_1$, and $\Gamma_1$ of the $C_s$ symmetry. In the other path, *Γ-P* line, the three-fold degenerate bands split into one two-fold degenerate band and one singly-degenerate band. Their irreducible representations are $\Gamma_3$ (2) and $\Gamma_1$ (1) of the $C_{3v}$ symmetry. It worth pointing out that it shows a quadratic dispersion along Γ-N and Γ-P paths, consistent with its 3D band structure in Fig. 2 (d). To characterize the nature of this quadratic three-fold excitation, we also establish an effective Hamiltonian via symmetry analysis. For $\Gamma$ point, its little group $T_d$ has four generators $\{C_{3,111}^-|000\}$, $\{C_{2z}|\frac{1}{2}0\frac{1}{2}\}$, $\{C_{2x}|\frac{1}{2}\frac{1}{2}0\}$ and $\{\widetilde{M}_{110}|\frac{1}{2}00\}$, and the irreducible representation at $\Gamma$ is chosen as $\Gamma_5$. Together with time reversal symmetry, the effective Hamiltonian constrained by symmetries is expressed as

$$\mathcal{H}(\mathbf{k}) = \begin{pmatrix} \alpha(k_x^2 + k_y^2) & i(\beta_- k_+^2 + \beta_+ k_-^2) & -\gamma k_- k_z \\ -i(\beta_+ k_+^2 + \beta_- k_-^2) & \alpha(k_x^2 + k_y^2) & i\gamma k_+ k_z \\ -\gamma_+ k_z & -i\gamma k_- k_z & 2\alpha k_z^2 \end{pmatrix}, \quad (3)$$

with $k_\pm = k_x \pm ik_y$, and α, γ are real parameters, $\beta_\pm = (2\alpha \pm \sqrt{2}\gamma)/4$. Seeing from this effective Hamiltonian, on plane $k_z = 0$, this model could resemble a 2D double Weyl point whose Chern number is determined by $2\text{sgn}(|\beta_+| - |\beta_-|)$. To derive clear surface states, we make a projection on plane (0 1 0).

Lastly, we discuss the four-fold point at *P* point. As shown in Fig. 2(b), the four-fold degenerate point is a result of the crossing of two non-degenerate bands and one double-degenerate band along path *Γ-P-H*. In details, the double degenerate band and two singly-degenerate bands separately belong to irreducible representations $\Gamma_3$ (2), $\Gamma_1$ (1) and $\Gamma_2$ (1) of the $C_{3v}$ symmetry. Notably, the 3D band structure (see in Fig. 2(e)) for P point indicates that it has a linear dispersion. To confirm its topology, we construct an effective model around the band crossings at P. There are four independent generators at P point, namely, $\{C_{3,\overline{1}11}^+|0\frac{1}{2}\frac{1}{2}\}$, $\{C_{2y}|0\frac{1}{2}\frac{1}{2}\}$, $\{C_{2x}|\frac{3}{2}\frac{3}{2}0\}$ and

$\{S_{4x}^+|\frac{1}{2}11\}$. The irreducible representation for the four-fold degenerate band at $P$ is $\Gamma_3$. With the constrains from symmetries, we expand the model up to the first order which is given as

$$\mathcal{H}_{\text{Dirac}}(\boldsymbol{k}) = \begin{pmatrix} 0 & Ck_- & -Ae^{i\theta_1}k_z & Be^{-i\theta_2}k_+ \\ Ck_+ & 0 & Be^{-i\theta_2}k_- & Ae^{i\theta_1}k_z \\ -Ae^{-i\theta_1}k_z & Be^{i\theta_2}k_+ & 0 & Ck_- \\ Be^{i\theta_2}k_- & Ae^{-i\theta_1}k_z & Ck_+ & 0 \end{pmatrix} \quad (4)$$

Here, we should point out that parameters A and B are the functions of C, and $\theta_1, \theta_2$ are real parameters. Consider a limitation $C \to 0$, the model can be given by

$$\mathcal{H}_{\text{Dirac}}(\boldsymbol{k}) = \begin{pmatrix} 0 & 0 & Dk_z & -Dk_+ \\ 0 & 0 & -Dk_- & -Dk_z \\ Dk_z & -Dk_+ & 0 & 0 \\ -Dk_- & -Dk_z & 0 & 0 \end{pmatrix}, \quad (5)$$

with D is a real parameter. Such a model indicates that this Dirac point is a composition of two Weyl points of the same chirality. Hence, this Dirac point is a charge-2 Dirac point, with $\mathcal{C} = \pm 2$. At this point, one has checked that the Chern number equal to $-2$.

We have demonstrated that the different excitations are stabilized by the crystalline symmetries. The most fascinating discovery is that the six-fold excitation carrying a zero topological charge which can be regarded as a composition of two spin-1 fermions. And the fourfold point carries a nonzero Chern number. Next, we focus on the topological surface Fermi arcs for them.

## B. Surface states

A nontrivial topological phase implies the presence of surface states. We plot the surface spectrum for $Yb_2C_3$ on the plane (0 1 0) [see in Fig. 3(a)], in which two Fermi arcs are emanating from the projections (*i.e.*, $\bar{P}, \bar{H}$, and $\bar{\Gamma}$) of the band crossings, including six-fold, four-fold and three-fold points. The enlarged images for surface spectra are given in Fig. 3(b-d) on different surfaces, one can observe that there are indeed two Fermi arcs emanating from the corresponding projections. The clear Fermi arc states on different surfaces can greatly facilitate their detections in future experiments. Most interestingly, we find that the surface states are extended in the whole BZ. The reason for it may be that these fermionic excitations reside in either the center or the corner of the BZ, the surface states that connect their projections emerge extensively on the side surface. In addition, we also calculated the constant energy slices near $\bar{P}, \bar{H}$ and $\bar{\Gamma}$ corresponding to Fig. 3(a) (at $E_1$ = 0.16 eV, $E_2 = E_F$, and $E_3$ = -0.08 eV), as shown in Fig. 3(e). In these energy levels, the profile for the Fermi arcs can also be identified.

### C. Effects of strain

Based on the slope of the crossing bands, a nodal point can be classified into type-I and type-II classes. On the other hand, they also can be classified into distinct categories according to their energy dispersions around the band crossings, namely linear, quadratic and cubic Weyl points. For nodal lines, depending on the type of nodal points on them, it is proposed that type-I, type-II, and hybrid-type nodal lines. When all the points on the nodal line are type-I (type-II), the nodal line corresponds to type-I (type-II); if the nodal line possesses both type-I and type-II nodal points, it belongs to the hybrid type. Many interesting properties have been proposed for each type of nodal points and nodal lines. In our work, we can realize these distinct topological phases by symmetry breaking via strain. Here, we take $Yb_2C_3$ as an example.

First, we use a compressive strain along [001] to break the threefold rotation along [111] direction. One can see that it makes the six-fold point split into three two-fold degenerate points at $H$ point due to the breaking of $C_{3,111}$. These three points are not isolated, but points belonging to nodal lines on path $H$-$N$. Meanwhile, there emerges a type-II Weyl point on path $\Gamma$-$H$. Turn to three-fold point at $\Gamma$ point, there is a single band keeping away from the degenerate point, and a type-I quadratic Weyl point appears at $\Gamma$ point which almost lies at the Fermi level [see Fig. 4 (e)]. A type-II Weyl point also emerges on the $\Gamma$-$N$ path. In Fig. 4 (h), under a [0 0 1] direction strain, the four-fold point is separated into two double degenerate points at $P$ point.

Second, we consider a strain along [1 1 1] direction. Then, the six-fold point at H point is transformed into a four-fold point and a two-fold degenerate point, and the two-fold point is also a point on the nodal line along path *H-N* [see in Fig. 4(c)]. Such a strain makes the three-fold point at Γ point split into one two-fold degenerate point and one singly-degenerate band [see Fig. 4 (f)]. Besides, there are also emerging a type-I Weyl point along the *N-Γ* path and a three-fold point along *Γ-P* line. The four-fold point is transformed into one double degenerate point which is a point on nodal line on path P-Γ and two singly-degenerate bands [see Fig. 4(i)]. Moreover, there emerge three-fold points on path *Γ-P*, togethering with a nodal line along Γ-P and a nodal loop around *P* point. Notably, for each *k*-path (*P-a*, *P-b*, *P-c* and *P-d*) from the *P* point in the plane, we can obtain both type-I and type-II band crossings, as shown in Fig. 4(j). It is evident that the nodal loop in Fig. 4(i) is a hybrid nodal loop. Here, we should address the difference between three-fold points along *Γ-P* and the one appearing at *Γ* point model: First, the former one appearing on high symmetry path is the crossing between a 2D and a 1D irreducible representations. However, the latter occurs at a high symmetry point which is in a 3D irreducible representation. Second, the former one is accompanied by nodal lines, the topology is strongly related to the Berry phase of nodal lines [55]. The topology of the latter one is exhibited by its topological charge and corresponding surface states.

Beyond the strain, when SOC is taken into consideration, the topological phases would experience a transformation. We discuss the case with SOC in the Supporting Information (SI). We also find the presence of some unconventional fermions with SOC.

## V. DISCUSSION AND CONCLUSION

Before closing, we have several remarks. First, the most important finding here is that we propose $Yb_2C_3$ as an ideal candidate to study multiple types of unconventional fermions, including the six-, four- and three-fold fermions. Because the band crossings in the $Yb_2C_3$ electric band structure are closed to the Fermi level, and it has a "clean" band structure without other rambling bands nearby. Besides, in this work, we find that there are also other twelve existing materials, namely $Re_2C_3$ (Re = Y, La, Ce, Pr, Nd, Sm, Tb, Dy, Ho, Er, Tm, Lu), possessing the six-fold, three-fold and four-fold points. Their band structures (see Fig. 5) are similar to that of $Yb_2C_3$. In stark contrast to Weyl point with $\pm 1$ topological charge and Dirac point with zero topological charge, there is a charge-2 Dirac points in $Yb_2C_3$. Therefore, some physical phenomena about topological charges are likely to be observed.

Second, the $Yb_2C_3$ has clear "Fermi arcs" connected these unconventional fermions. Because these fermionic excitations reside in either the center or the corner of the BZ, the surface states that connect their projections emerge extensively on the side surface. Moreover, these Fermi arcs are within wide energy window. Thus, it is easy to observe experimentally.

Third, the six-, four- and three-fold points in $Yb_2C_3$ also show two additional features: (i) The strain induces various topological phases in $Yb_2C_3$. Under a [0 0 1] direction strain, these points can be transformed to type-I quadratic and type-II linear Weyl points. Under a [1 1 1] direction strain, these points could be transformed to four-fold, three-fold, type-I Weyl points and hybrid type nodal line. (ii) When SOC is considered, there exist four-fold and eight-fold (higher degeneracy) band crossings [see Figure S2 in SI]. It indicates that the study still is meaningful after considered SOC.

In conclusion, by using first-principles calculations, we predict that bulk $Re_2C_3$ family compounds host six-fold, three-fold and four-fold fermions in the absence of SOC. Even when the *U* value varies from 0 eV to 5 eV (see in SI), the multifold degeneracy points are still stable. Most importantly, $Yb_2C_3$ is an ideal material for studying unconventional quasiparticles in conventional crystals for its advantages over other materials. Thereby, $Re_2C_3$ family compounds, especially $Yb_2C_3$, can serve as new platforms for the study of unconventional quasiparticles in the near future experiments.

**Acknowledgements**

This work is supported by National Natural Science Foundation of China (Grants No. 12004096), "100 Talents Plan" of Hebei Province (No. E2020050014), "State Key

Laboratory of Reliability and Intelligence of Electrical Equipment" of Hebei University of Technology (No. EERI_PI2020009), the Nature Science Foundation of Hebei Province (Grants No. E2019202222 and No. E2019202107) and Doctoral Postgraduate Innovation Funding project of Hebei Province (No. CXZZBS2021028). One of the authors (X.M. Zhang) acknowledges the financial support from Young Elite Scientists Sponsorship Program by Tianjin.

# References

[1] P. B. Pal, Am. J. Phys. **79**, 485 (2011).

[2] X. Wan, A. M. Turner, A. Vishwanath, and S. Y. Savrasov, Phys. Rev. B **83**, 205101 (2011).

[3] H. Weng, C. Fang, Z. Fang, B. A. Bernevig, and X. Dai, Phys. Rev. X **5**, 011029 (2015).

[4] B. Q. Lv, H. M. Weng, B. B. Fu, X. P. Wang, H. Miao, J. Ma, P. Richard, X. C. Huang, L. X. Zhao, G. F. Chen, Z. Fang, X. Dai, T. Qian, and H. Ding, Phys. Rev. X **5**, 031013 (2015).

[5] Y. Xu, F. Zhang, and C. Zhang, Phys. Rev. Lett. **115**, 265304 (2015).

[6] Z. Wang, D. Gresch, A. A. Soluyanov, W. Xie, S. Kushwaha, X. Dai, M. Troyer, R. J. Cava, and B. A. Bernevig, Phys. Rev. Lett. **117**, 056805 (2016).

[7] G. Autès, D. Gresch, M. Troyer, A. A. Soluyanov, and O. V. Yazyev, Phys. Rev. Lett. **117**, 066402 (2016).

[8] G. Chang *et al.*, Sci. Adv. **2**, e1600295 (2016).

[9] J. Ruan, S. K. Jian, H. Yao, H. Zhang, S. C. Zhang, and D. Xing, Nat. Commun. **7**, 11136 (2016).

[10] J. Ruan, S. K. Jian, D. Zhang, H. Yao, H. Zhang, S.-C. Zhang, and D. Xing, Phys. Rev. Lett. **116**, 226801 (2016).

[11] W. Z. Meng, X. M. Zhang, T. L. He, L. Jin, X. F. Dai, Y. Liu, and G. D. Liu, J. Adv. Res. **24**, 523–528 (2020).

[12] G. Sharma, P. Goswami, and S. Tewari, Physical Review B **96**, 045112 (2017).

[13] S. M. Young, S. Zaheer, J. C. Y. Teo, C. L. Kane, E. J. Mele, and A. M. Rappe, Phys. Rev. Lett. **108**, 140405 (2012).

[14] Z. Wang, H. Weng, Q. Wu, X. Dai, and Z. Fang, Phys. Rev. B **88**, 125427 (2013).

[15] Y. X. Zhao and Z. D. Wang, Phys. Rev. Lett. **110**, 240404 (2013).

[16] Z. K. Liu, B. Zhou, Y. Zhang, Z. J. Wang, H. M. Weng, D. Prabhakaran, S.-K. Mo, Z. X. Shen, Z. Fang, X. Dai, Z. Hussain, and Y. L. Chen, Science **343**, 864 (2014).

[17] M. Neupane, S.-Y. Xu, R. Sankar, N. Alidoust, G. Bian, C. Liu, I. Belopolski, T.-R. Chang, H.-T. Jeng, H. Lin, A. Bansil, F. C. Chou, and M. Z. Hasan, Nat. Commun. **5**, 3786 (2014).

[18] Z. K. Liu, J. Jiang, B. Zhou, Z. J. Wang, Y. Zhang, H. M. Weng, D. Prabhakaran, S.-K. Mo, H. Peng, P. Dudin, T. Kim, M. Hoesch, Z. Fang, X. Dai, Z. X. Shen, D. L. Feng, Z. Hussain, and Y. L. Chen, Nat. Mater. **13**, 677 (2014).

[19] S.-Y. Xu, C. Liu, S. K. Kushwaha, R. Sankar, J. W. Krizan, I. Belopolski, M. Neupane, G. Bian, N. Alidoust, T.-R. Chang, H.-T. Jeng, C.-Y. Huang, W.-F. Tsai, H. Lin, P. P. Shibayev, F.-C. Chou, R. J. Cava, and M. Z. Hasan, Science **347**, 294 (2015).

[20] Y. Du, F. Tang, D. Wang, L. Sheng, E.-J. Kan, C.-G. Duan, S. Y. Savrasov, and X. Wan, npj Quantum Mater. **2**, 3 (2017).

[21] F. Tang, H. C. Po, A. Vishwanath, and X. Wan, Nature (London) **566**, 486 (2019).

[22] T. Zhang, Y. Jiang, Z. Song, H. Huang, Y. He, Z. Fang, H. Weng, and C. Fang, Nature (London) **566**, 475 (2019).

[23] B. Bradlyn, J. Cano, Z. Wang, M. Vergniory, C. Felser, R. Cava, and B. A. Bernevig, Science **353**, aaf5037 (2016).

[24] B. J. Wieder, Y. Kim, A. M. Rappe, and C. L. Kane, Phys. Rev. Lett. **116**, 186402 (2016).

[25] H. Weng, C. Fang, Z. Fang, and X. Dai, Phys. Rev. B **94**, 165201 (2016).

[26] H. Weng, C. Fang, Z. Fang, and X. Dai, Phys. Rev. B **93**, 241202 (2016).

[27] Z. Zhu, G. W. Winkler, Q. S. Wu, J. Li, and A. A. Soluyanov, Phys. Rev. X **6**, 031003 (2016).

[28] B. Lv *et al*., Nature (London) **546**, 627 (2017).

[29] P. Tang, Q. Zhou, and S.-C. Zhang, Phys. Rev. Lett. **119**, 206402 (2017).

[30] D. Takane, Z. Wang, S. Souma, K. Nakayama, T. Nakamura, H. Oinuma, Y. Nakata, H. Iwasawa, C. Cacho, T. Kim, K. Horiba, H. Kumigashira, T. Takahashi, Y. Ando, and T. Sato, Phys. Rev. Lett. **122**, 076402 (2019).

[31] T. Zhang, Z. Song, A. Alexandradinata, H. Weng, C. Fang, L. Lu, and Z. Fang, Phys. Rev. Lett. **120**, 016401 (2018).

[32] Z. Rao, H. Li, T. Zhang, S. Tian, C. Li, B. Fu, C. Tang, L. Wang, Z. Li, W. Fan, J. Li, Y. Huang, Z. Liu, Y. Long, C. Fang, H. Weng, Y. Shi, H. Lei, Y. Sun, T. Qian, and H. Ding, Nature 567, 28 (2019).

[33] G. Chang, S.-Y. Xu, B. J. Wieder, D. S. Sanchez, S.-M. Huang, I. Belopolski, T.-R. Chang, S. Zhang, A. Bansil, H. Lin, and M. Z. Hasan, Phys. Rev. Lett. **119**, 206401 (2017).

[34] F. Flicker, F. de Juan, B. Bradlyn, T. Morimoto, M. G. Vergniory, and A. G. Grushin, Phys. Rev. B **98**, 155145 (2018).

[35] F. de Juan, A. G. Grushin, T. Morimoto, and J. E. Moore, Nat. Commun. **8**, 15995 (2017).

[36] N. B. M. Schröter, D. Pei, M. G. Vergniory, Y. Sun, K. Manna, F. D. Juan, J. A. Krieger, V. Süss, M. Schmidt, P. Dudin, B. Bradlyn, T. K. Kim, T. Schmitt, C. Cacho, C. Felser, V. N. Strocov, and Y. Chen, Nat. Phys. **15**, 759–765(2019).

[37] Z. P. Sun, C. Q. Hua, X. L. Liu, Z. T. Liu, M. Ye, S. Qiao, Z. H. Liu, J. S. Liu, Y. F. Guo, Y. H. Lu, and D. W. Shen, Phys. Rev. B **101**, 155114 (2020).

[38] X. Yáng, T. A. Cochran, R. Chapai, D. Tristant, J.-X. Yin, I. Belopolski, Z. Chéng, D. Multer, S. S. Zhang, N. Shumiya, M. Litskevich, Y. Jiang, G. Chang, Q. Zhang, I. Vekhter, W. A. Shelton, R. Jin, S.-Y. Xu, and M. Z. Hasan, Phys. Rev. B **101**, 201105(R) (2020).

[39] N. Kumar, M. Yao, J. Nayak, M. G. Vergniory, J. Bannies, Z. Wang, N. B. M. Schröter, V. N. Strocov, L. Müchler, W. Shi, E. D. L. Rienks, J. L. Mañes, C. Shekhar, S. S. P. Parkin, J. Fink, G. H. Fecher, Y. Sun, B. A. Bernevig, and C. Felser, Adv.

Mater. 1906046 (2020).

[40] S. Thirupathaiah, Y. S. Kushnirenko, K. Koepernik, B. R. Piening, B. Büchner, S. Aswartham, J. V. D. Brink, S. V. Borisenko, I. C. Fulga, arXiv:2006.08642v1.

[41] S. Nie, B. A. Bernevig, and Z. Wang, arXiv:2006.12502v1.

[42] X. Zhang, Q. Gu, H. Sun, T. Luo, Y. Liu, Y. Chen, Z. Shao, Z. Zhang, S. Li, Y. Sun, Y. Li, X. Li, S. Xue, J. Ge, Y. Xing, R. Comin, Z. Zhu, P. Gao, B. Yan, J. Feng, M. Pan, and J. Wang, Phys. Rev. B **102**, 035125 (2020).

[43] G. Kresse and D. Joubert, Phys. Rev. B **59**, 1758 (1999).

[44] P. E. Blöchl, Phys. Rev. B **50**, 17953 (1994).

[45] J. P. Perdew, K. Burke, and M. Ernzerhof, Phys. Rev. Lett. **77**, 3865 (1996).

[46] V. I. Anisimov, J. Zaanen, and O. K. Andersen, Phys. Rev. B 44, 943 (1991).

[47] N. Marzari and D. Vanderbilt, Phys. Rev. B: Condens. Matter Mater. Phys. **56**, 12847 (1997).

[48] A. A. Mostofi, J. R. Yates, Y.-S. Lee, I. Souza, D. Vanderbilt, and N. Marzari, Comput. Phys. Commun. 178, 685 (2008).

[49] Q. S. Wu, S. N. Zhang, H.-F. Song, M. Troyer, and A. A. Soluyanov, Comput. Phys. Commun. 224, 405 (2018).

[50] V. I. Novokshonov, Zhurnal Neorganicheskoi Khimii 25, 684-689 (1980).

[51] M. Atoji and D. E. Williams, J. Chem. Phys. 35(6), 1960-1966 (1961).

[52] M. Atoji, J. Solid. State. Chem. **26**, 51-57 (1978).

[53] F. H. Spedding, K. A. J. Gschneidner, and A. H. Daane, J. Am. Chem. Soc. 80, 4499-4503 (1958).

[54] M. Atoji and Y. Tsunoda, J. Chem. Phys. 54, 3510-3513 (1971).

[55] G. W. Winkler, S. Singh, and A. A. Soluyanov, Chin. Phys. B **28**, 7 (2019).

# Figures and captions:

Table 1 The optimized and experimental lattice parameters and the energy positions of multifold degeneracy points (Unit: eV) for $Re_2C_3$ (Re = Y, La, Ce, Pr, Nd, Sm, Tb, Dy, Ho, Er, Tm, Yb, Lu) compounds. Here OL and EL stand for optimized and experimental lattices. SP, TP and FP stand for the six-fold, three-fold and four-fold points, respectively.

| Compound | OL (Å) | EL (Å) | Ref. | SP (eV) | TP (eV) | FP (eV) |
|---|---|---|---|---|---|---|
| $Y_2C_3$ | 8.264 | 8.234 | 50 | -0.773 | -0.998 | -0.781 |
| $La_2C_3$ | 8.819 | 8.817 | 51 | -0.559 | -1.013 | -0.614 |
| $Ce_2C_3$ | 8.405 | 8.448 | 51 | -0.661 | -1.281 | -0.751 |
| $Pr_2C_3$ | 8.700 | 8.590 | 52 | -0.666 | -1.021 | -0.662 |
| $Nd_2C_3$ | 8.617 | 8.534 | 52 | -0.683 | -1.033 | -0.681 |
| $Sm_2C_3$ | 8.477 | 8.399 | 53 | -0.714 | -1.043 | -0.716 |
| $Tb_2C_3$ | 8.289 | 8.253 | 51 | -0.762 | -1.057 | -0.775 |
| $Dy_2C_3$ | 8.237 | 8.206 | 52 | -0.778 | -1.062 | -0.798 |
| $Ho_2C_3$ | 8.189 | 8.144 | 54 | -0.793 | -1.067 | -0.820 |
| $Er_2C_3$ | 8.143 | 8.132 | 50 | -0.809 | -1.072 | -0.843 |
| $Tm_2C_3$ | 8.090 | 8.092 | 50 | -0.828 | -1.082 | -0.874 |
| $Yb_2C_3$ | 8.348 | 8.072 | 50 | 0.194 | -0.004 | -0.011 |
| $Lu_2C_3$ | 8.011 | 8.035 | 50 | -0.856 | -1.079 | -0.912 |

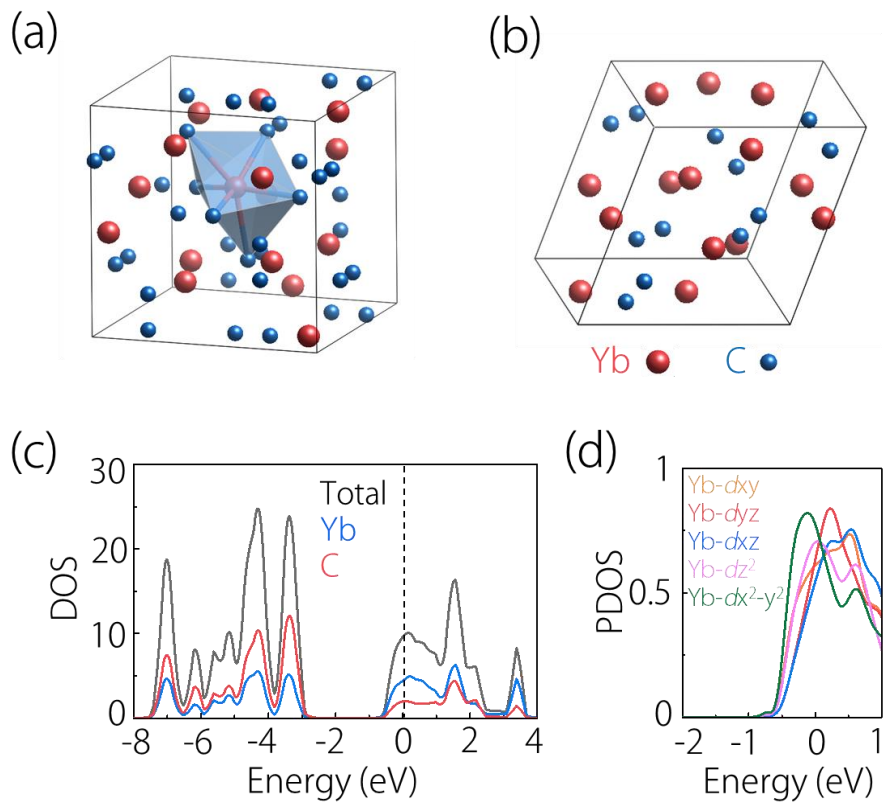

Fig. 1 (a) Unit cell and (b) primitive cell structure for $Yb_2C_3$. The shadowed region in (a) shows that one Yb atom is bonded to six C atoms. (c) The total density of state (DOS). (d) The projected density of state (PDOS).

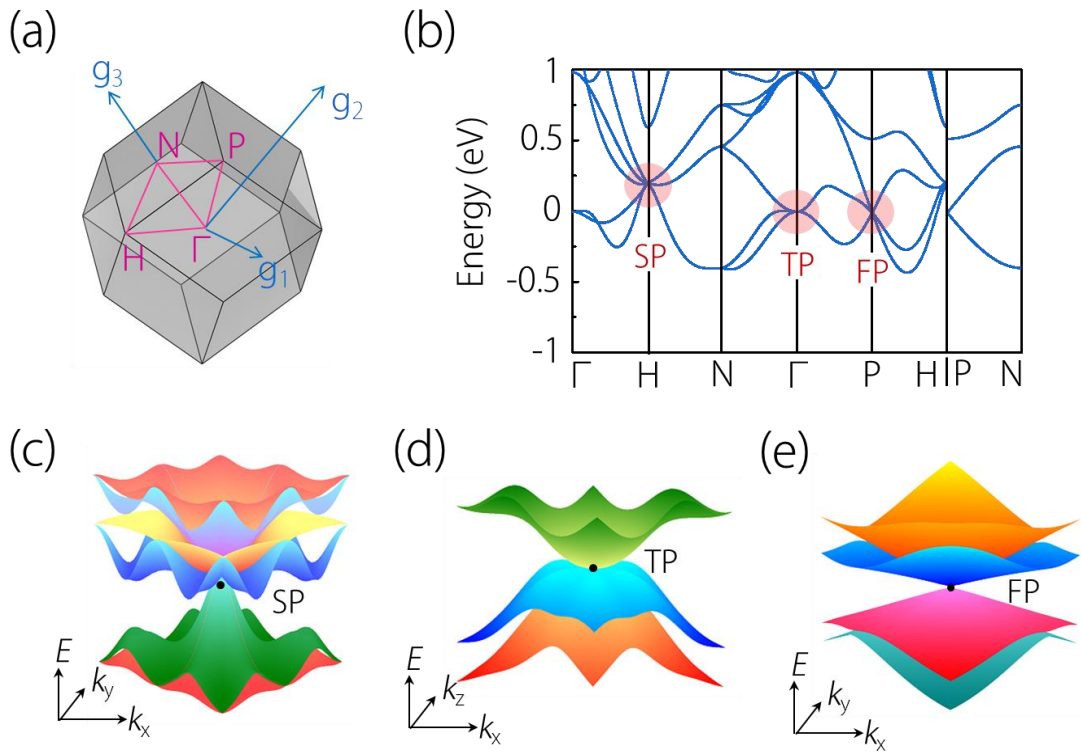

Fig. 2 (a) The bulk Brillouin zone for Yb$_2$C$_3$. (b) Electronic band structure for Yb$_2$C$_3$ without SOC. The shadowed regions show the positions of six-fold, three-fold and four-fold points, which are labeled as SP, TP and FP, respectively. (c), (d) and (e) are shown the three-dimensional plots of band dispersions near SP, TP and FP.

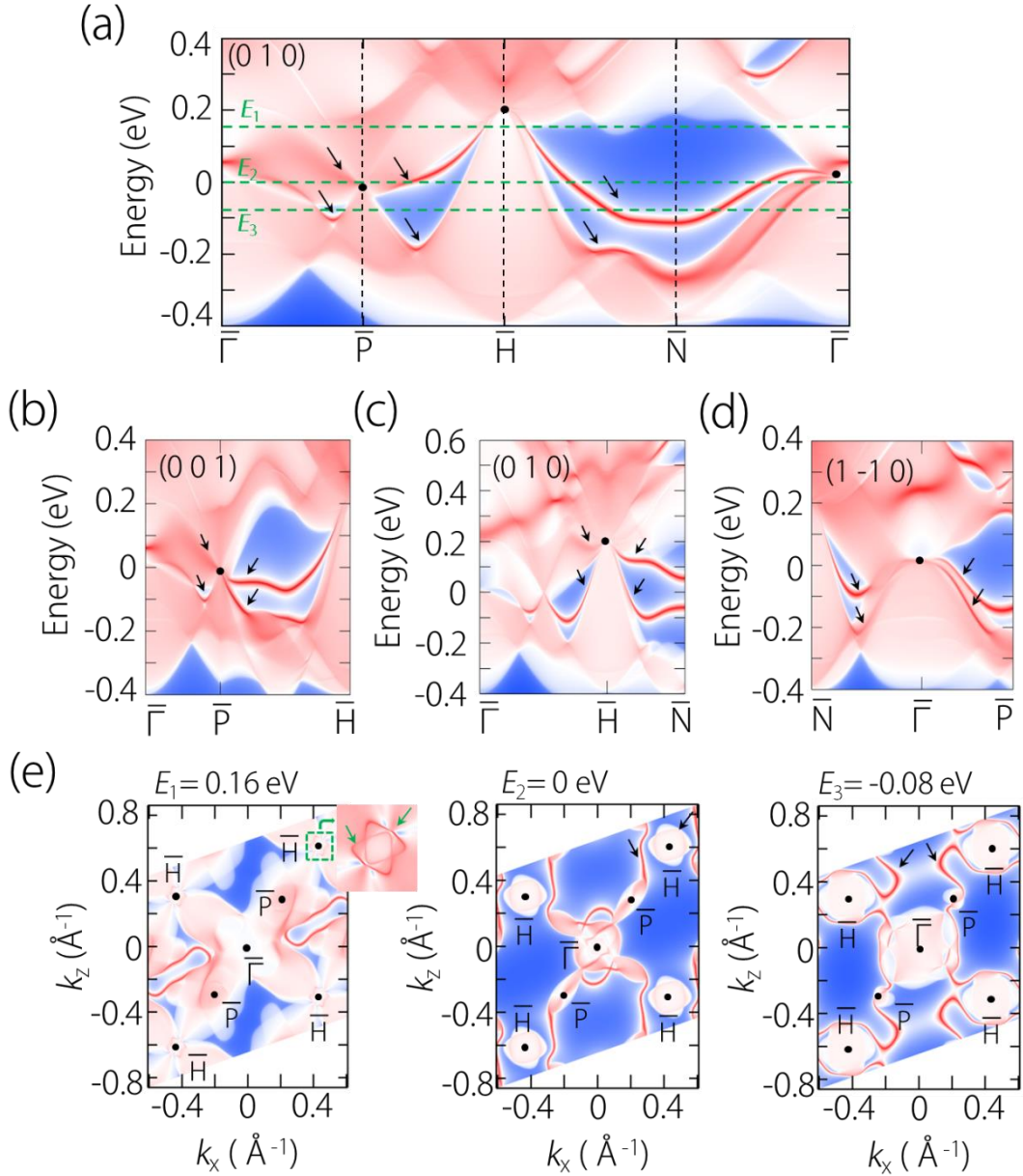

Fig. 3 (a) Projected spectrum on the (0 1 0) surface of $Yb_2C_3$. The black dot indicates the positions of the four-fold, six-fold and three-fold points. Two Fermi arcs (pointed by arrows) connect these multifold-degenerate points. (b), (c) and (d) Surface states of other projective or opposite terminating surfaces with clearer Fermi arcs. (e) The corresponding constant energy slice for (a) at $E_1$ = 0.16 eV, $E_2$ = 0 eV and $E_3$ = -0.08 eV. The arrows point to Fermi arcs.

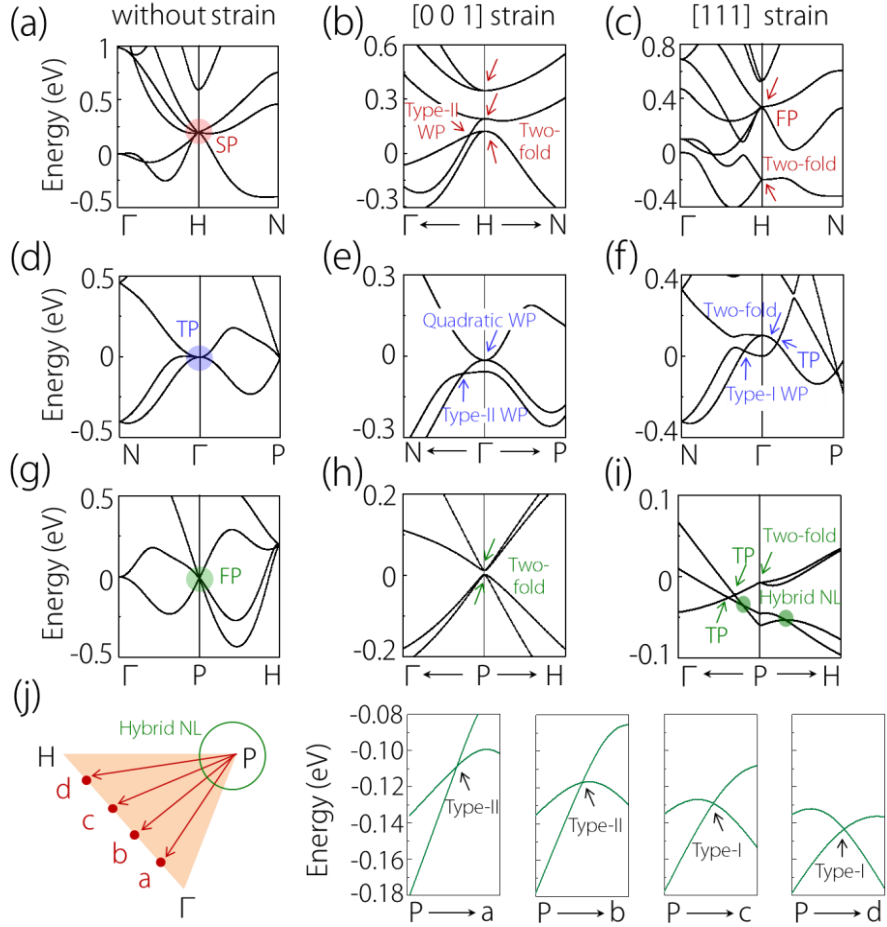

Fig. 4 Bulk band structures along *Γ-H-N* path without strain (a), with [0 0 1] strain (b), and with [1 1 1] strain (c). Bulk band structures along *N-Γ-P* path without strain (d), with [0 0 1] strain (e), and with [1 1 1] strain (f). Bulk band structures along *Γ-P-H* path without strain (g), with [0 0 1] strain (h), and with [1 1 1] strain (i). (j) Schematic illustration of the hybrid type nodal line in (i) centering on the *P* point in the *H-P-Γ* plane. Crossing the nodal line, we choose four *k*-paths, namely *P-a*, *P-b*, *P-c* and *P-d*. The points *a*, *b*, *c* and *d* are equally spaced between *Γ* and *H*. Electronic band structures of $Yb_2C_3$ at the *P-a*, *P-b*, *P-c* and *P-d* paths display that nodal line possesses both type-I and type-II band crossings. SP: six-fold point; FP: four-fold point; TP: three-fold point; WP: Weyl point; NL: nodal line.

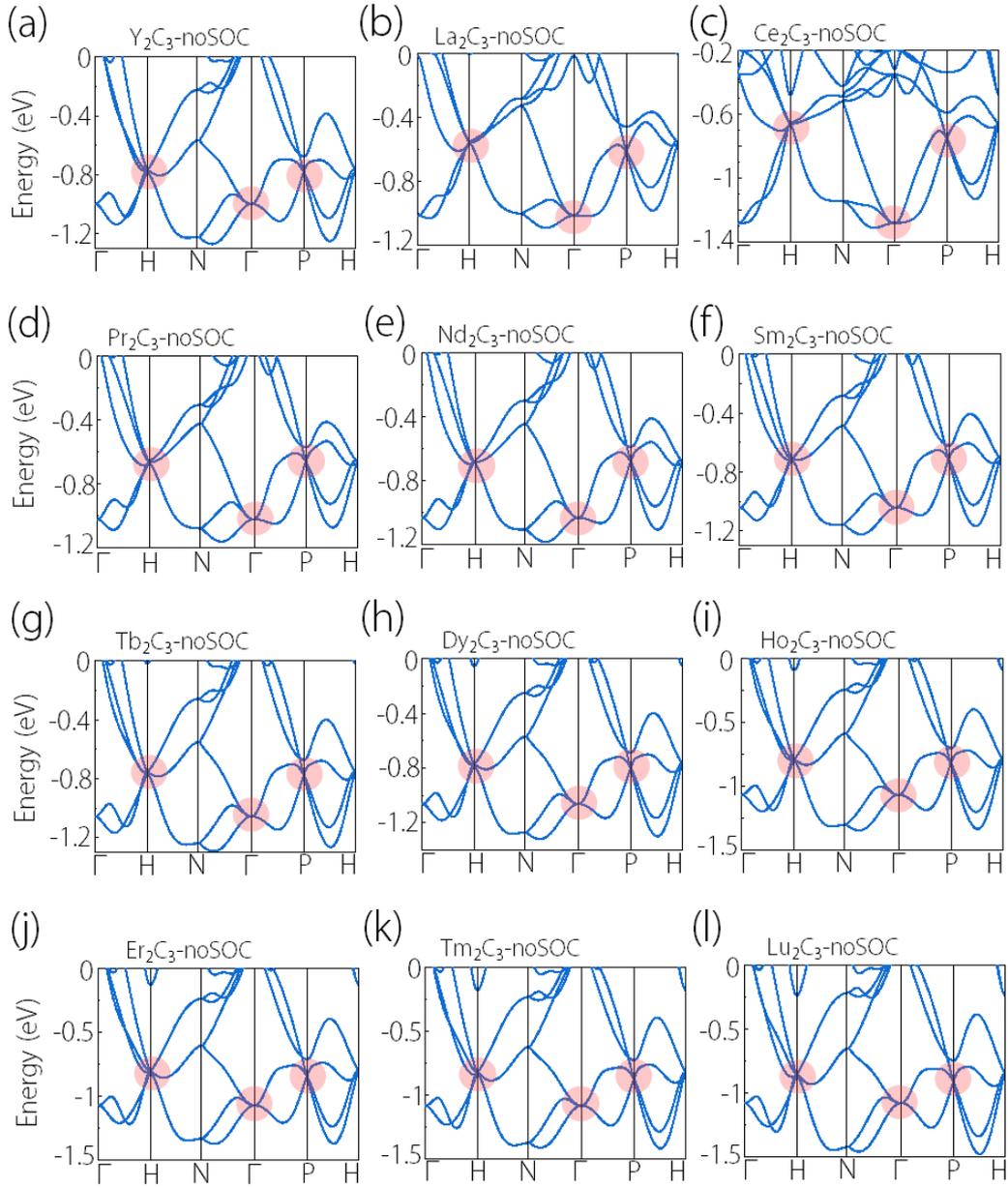

Fig. 5 Electronic band structures of (a) $Y_2C_3$, (b) $La_2C_3$, (c) $Ce_2C_3$, (d) $Pr_2C_3$, (e) $Nd_2C_3$, (f) $Sm_2C_3$, (g) $Tb_2C_3$, (h) $Dy_2C_3$, (i) $Ho_2C_3$, (j) $Er_2C_3$, (k) $Tm_2C_3$ and (l) $Lu_2C_3$ without SOC. The multifold degeneracy points are highlighted by red shadowed regions. Six-fold point is at $H$ point; Three-fold point is at $\Gamma$ point; Four-fold point is at $P$ point.